\documentclass[12pt,preprint]{aastex}

\slugcomment{}

\shorttitle{On the Pioneer Anomaly}
\shortauthors{Tren\v{c}evski \& Celakoska}

\begin{document}

\title{On the Pionner Anomaly}

\author{Kostadin Tren\v{c}evski}
\affil{Faculty of Natural Sciences and Math., Ss. Cyril and Methodius Univ.,
P.O.Box 162, 1000 Skopje, Macedonia}
\email{kostatre@iunona.pmf.ukim.edu.mk} 

\and 

\author{Emilija G. Celakoska}
\affil{Faculty of Mechanical Engineering, Ss. Cyril and Methodius Univ., 
P.O.Box 464, 1000 Skopje, Macedonia}
\email{cemil@ereb1.mf.ukim.edu.mk}

\begin{abstract}
In this paper an explanation of the Pioneer anomaly is given, 
using the hypothesis of the time dependent 
gravitational potential. The 
implications of this hypothesis on the planetary orbits and orbital periods
are given in section 2. 
In sections 3 and 4 we give a detailed explanation of the Pioneer anomaly, 
which corresponds to the observed phenomena. 
\end{abstract}


\section{Introduction} 

In 1998 an interesting paper by Anderson et al. (1998)
was published, with indication that 
there is an unexplained frequency shift in the navigation of the spacecraft 
Pioneer 10 and 11. Namely, the detailed analysis of the frequencies 
of the round-trip signals from the 
spacecraft Pioneer 10 and Pioneer 11, yield to a slight disagreement with 
the expected frequencies. This small shift of the frequencies accumulates 
in time. Later a very detailed analysis of this phenomena
was given by Anderson et al.(2002). 
Firstly, we give a short history about this effect, following the 
references (Anderson et al. 1998, 2002; Turyshev et al. 2004, 2005, 2006). 

The spacecraft Pioneer 10 was launched on 1972 March 2, while Pioneer 11 
was launched on 1973 April 5, and their aim was to explore the outer Solar 
System (Turyshev et al. 2006). 
They are the first spacecrafts that left the Solar System. 
The mentioned phenomena was detected in 1980 for the spacecraft 
Pioneer 10 for the first time, 
when it was about 20 AU far from the Sun, and the influence 
of the neighboring planets was negligible on the detected shift frequency. 
This phenomena was well examined for Pioneer 10 
on the heliocentric distances 20-70 AU. 
Similar results were also obtained for the spacecraft Pioneer 11, but, 
unfortunately, after 1995 November 
the contact with Pioneer 11 was lost, 
and the research with Pioneer 11 was stopped. 
The last signal from Pioneer 10 was obtained in April 2002, 
when it was about 80 AU from the Sun. 

The unexplained blue shift of the frequency is denoted by $a_P$ 
and it is introduced by the formula (15) in (Anderson et al. 2002), i.e. 
$$[\nu_{obs}(t)-\nu_{model}(t)]_{DSN} =-\nu_0\frac{2a_Pt}{c},
\qquad \nu_{model}=
\nu_0\Bigl [1-2\frac{v_{model}(t)}{c}\Bigr ],\eqno{(1.1)}$$
where $\nu_0$ is the reference frequency, the factor 2 is because of 
the two way 
data, $v_{model}$ is the modeled velocity of the spacecraft due to the 
gravitational and other forces and $\nu_{obs}$ is frequency of the 
re-transmitted signal observed by DSN antennae. 
Since $a_P$ is in the same units as the acceleration, it is also called 
anomaly acceleration and it is interpreted as if there is an acceleration 
toward the Sun of magnitude 
$a_P\approx (8.74\pm 1.33)\times 10^{-8}$cm s$^{-2}$. The existence of 
such an acceleration is almost impossible, because it would change the 
planetary orbits, while the astronomical observations do not show such 
change (Iorio 2006). 
So it would be more convenient to say 
"Pioneer anomaly" than "anomaly acceleration". 
It is mentioned in (Anderson et al. 1998, 2002) 
that without using 
the acceleration $a_P$, the anomalous frequency shift can be 
interpreted as a 
"clock acceleration" $-a_t=-2.8\times 10^{-18}$s s$^{-2}$. 

There are several attempts to explore the possible sources of the 
Pioneer anomaly. The original efforts, including classical 
explanations (effects from sources external to the spacecraft, 
study of the on-board systematics and computational systematics)
of this anomaly are not strong enough to explain the observed Doppler shift 
(Turyshev et al. 2006). 
So, the question of a possible new physics arises. 
There are several attempts to explore the possible sources of the 
Pioneer anomaly via non-classical Newtonian mechanics and general 
relativity. The ideas are quite different and here we mention some of them: 
Newtonian gravity as a long wavelength excitation of a scalar condensate 
inducing electroweak symmetry breaking (Consoli \& Siringo 1999),
time dependence of the gravitational constant $G$ (Sidharth 2000), 
higher order (or Yukawa-like) 
corrections to the Newtonian potential (Anderson et al. 1998),
a scalar-tensor extension 
to the standard gravitational model (Clachi Novati et al. 2000),
interaction of the spacecraft
with a long-range scalar field coupled to gravity 
(Mbelek \& Lachi\`{e}ze-Rey 1999; Cadoni 2004), a length 
or momentum scale-dependent cosmological term in the gravitational 
action functional (Modanese 1999), 
a local solar system curvature  for 
light-like geodesics arising from the cosmological expansion 
(Rosales \& S\'{a}nchez-Gomez 1999), 
influence of the cosmological constant at the scale of the Solar System 
(Nottale 2003), 
a 5-dimensional cosmological model with a variable extra 
dimensional scale-factor and static external space (Belayev 2001), 
a superstrong interaction 
of photons or massive bodies with the graviton background yielding a 
constant acceleration proportional to the Hubble constant (Ivanov 2001), 
using an expanded PPN formalism (Ostvang 2002), 
explanation using the 
Brans-Dicke theory of gravity (Capozzielo et al. 2001), 
explanation using braneworld theories (Bertolami \& P\'{a}ramos 2004), 
and so on (Foot \& Volkas 2001; Guruprasad 1999, 2000; Wood \& Moreau 2001; 
Crawford 1999; Ingersoll et al. 1999; Milgrom 2001; 
Munyaneza \& Viollier 1999; Ra\~nada 2005). 

In this paper, we give new arguments and we upgrade the papers 
(Tren\v{c}evski 2005a, 2006) 
concerning the hypothesis the time dependent gravitational potential. 
In section 2 we give some results about the time dependent gravitational 
potential and in section 3 we will return to the Pioneer anomaly. 
The mathematical calculations will be shortened, 
while the physical interpretation will be emphasized. 

\section{Time dependent gravitational potential} 
In the recent paper (Tren\v{c}evski 2005a) 
and also in (Tren\v{c}evski 2006), 
an idea about linear 
change of the gravitational potential in the 
universe was developed. This linear (or almost linear) change is probably a 
consequence of the change of the density of the dark matter in the universe, 
but we shall consider only the consequence of this phenomena, not its 
origin. It can be interpreted in this way: the whole universe is inside of a 
shielded laboratory which falls freely in a gravitational field. Then, for 
a small time interval, the change of the gravitational potential can be 
considered to be almost linear. For simpler interpretation of the 
induced effects, we shall consider two observers: observer A in the shielded
laboratory, where the gravitational potential changes linearly, and 
observer B outside the shielded laboratory, far from the gravitational 
field, where the gravitational potential is constant.  
Thus the proper time for the observer B is uniform, i.e. not accelerated. 

The coefficient of the linear change of the gravitational potential is 
uniquely determined, so the observer A measures the Hubble redshift: 
The light signal, 
which starts from a distant galaxy, comes to the observer A after 
a long period, when the time is much faster, and so A observes a redshift. 

If a light signal starts from a star with a frequency $\nu_0$, then, 
according to the general relativity, 
after a period $t=\frac{R}{c}$ its frequency will be 
$$\nu =\nu_0\Bigl ( 1+\frac{R}{c^3}\frac{\partial V}{\partial t}\Bigr ).$$
It is assumed here that the sign 
of the gravitational potential $V$ is such that 
the gravitational potential $V$ is larger near the massive bodies. 
In comparison with the Hubble law, 
$$\nu = \nu_0\Bigl (1-\frac{RH}{c}\Bigr ),$$
where $H$ is the Hubble constant, $H\approx 70$ km s$^{-1}$ Mpc$^{-1}$, 
we obtain that 
$$\frac{\partial V}{\partial t}=-c^2H\approx -2\times 10^3
{\hbox {cm}^2}\;{\hbox {s}^{-3}}.\eqno{(2.1)}$$
Notice that this linear change of the gravitational potential 
in the universe is the same as the 
change of the Earth's gravitational potential
in a lift which moves toward the high floors, with velocity 2 cm s$^{-1}$. 

Now, accepting this linear change given by (2.1), the simplest 
application is the explanation of the 
Hubble redshift. Thus, the main reason for the Hubble redshift 
is the linear change of the gravitational potential, 
while the Doppler effect has a minor role. 

Further, we shall consider the influence 
to the planetary orbits and orbital periods 
(Tren\v{c}evski 2005a, 2006). 

Let us denote by $X,Y,Z,T$ the coordinates in the shielded laboratory 
observed by the observer B, and let us 
denote by $x,y,z,t$ the coordinates observed by the same observer B 
but in his close neighborhood. 
This can be interpreted in the following way. Assume that in the 
shielded laboratory there is motion of a particle around a body 
under the gravitation. Then $X,Y,Z$ are functions of $T$ on the trajectory.
Assume that, simultaneously, the same happens out of the shielded 
laboratory in a close neighborhood of the observer $B$. Then, for this 
simultaneous trajectory, $x,y,z$ will be functions of $t$. The 
observer $B$ is in a good position to compare these two trajectories. 

The coordinates $x,y,z,t$ will be called "normed 
coordinates", because the time is uniform and the dimensions of objects 
in a neighborhood of $B$ are static. 
According to the general relativity (GR), up to the first 
postnewtonian approximation we accept that 
$$
dx=\Bigl (1+\frac{V}{c^2}\Bigr )^{-1}dX = \Bigl (1+tH\Bigr )dX,\eqno{(2.2)}
$$
$$
dy=\Bigl (1+\frac{V}{c^2}\Bigr )^{-1}dY = \Bigl (1+tH\Bigr )dY,\eqno{(2.3)}
$$
$$
dz=\Bigl (1+\frac{V}{c^2}\Bigr )^{-1}dZ = \Bigl (1+tH\Bigr )dZ,\eqno{(2.4)}
$$
$$
dt=\Bigl (1-\frac{V}{c^2}\Bigr )^{-1}dT = \Bigl (1-tH\Bigr )dT.\eqno{(2.5)}
$$

Assuming that (2.2), (2.3), (2.4), and (2.5) are satisfied, 
it is not necessary to speak about the 
time dependent gravitational potential. 
Indeed, if we accept (2.2), (2.3), (2.4), and (2.5) as axioms, 
and neglect $H^2$, then $tH$ 
can be replaced by $TH$, $(1+tH)^{-1}=1-tH$, and so on. 
Further on, the Hubble redshift is a direct consequence from (2.5). 

{From} (2.2), (2.3), (2.4), and (2.5) we obtain 
$$
\Bigl (\frac{dX}{dT},\frac{dY}{dT},\frac{dZ}{dT}\Bigr )=
\Bigl (\frac{dx}{dt},\frac{dy}{dt},\frac{dz}{dt}\Bigr )(1-2tH)\eqno{(2.6)} 
$$
and by differentiating this equality by $T$ we get 
$$
\Bigl (\frac{d^2X}{dT^2},\frac{d^2Y}{dT^2},\frac{d^2Z}{dT^2}\Bigr )=$$
$$=\Bigl (\frac{d^2x}{dt^2},\frac{d^2y}{dt^2},\frac{d^2z}{dt^2}\Bigr ) 
-3tH \Bigl (\frac{d^2x}{dt^2},\frac{d^2y}{dt^2},\frac{d^2z}{dt^2}\Bigr )
-2H\Bigl (\frac{dX}{dT},\frac{dY}{dT},\frac{dZ}{dT}\Bigr ).\eqno{(2.7)} 
$$ 
In normed coordinates $x,y,z,t$ there is no acceleration caused by the 
time dependent gravitational potential, i.e. we can put there $H=0$. 
Thus, according to the coordinates $X,Y,Z,T$ there is an additional 
acceleration 
$$
-3tH \Bigl (\frac{d^2x}{dt^2},\frac{d^2y}{dt^2},\frac{d^2z}{dt^2}\Bigr )
-2\Bigl (H\frac{dX}{dT},H\frac{dY}{dT},H\frac{dZ}{dT}\Bigr ).
$$

It is very important to notice that the right sides of (2.2), (2.3) and (2.4)
are not total differentials, so there does not exist a 
functional dependence between the "coordinates" $x,y,z,t$ and $X,Y,Z,T$. 
This shows that we deal with unholonomic coordinates. 
The required functional dependence between $x,y,z,t$ and $X,Y,Z,T$ can exist 
only along a chosen curve, and it will depend on the choice of the curve. 
This yields to a violation of the strong equivalence principle, 
i.e. to one part of it, but on the other hand 
we shall see that the previous equations (2.2-5) yield to 
results compatible with the experimental measurements. The violation 
of the strong equivalence principle is explained in more details in 
(Tren\v{c}evski 2005b). 

In order to see the influence 
of the constant $H$ to the planetary orbits, we are looking for a 
functional dependence of the form 
$$ 
x=(1+\lambda tH)X,\; y=(1+\lambda tH)Y,\; z=(1+\lambda tH)Z,\; 
dt= (1-\mu tH)dT,\eqno{(2.8)}
$$
$\lambda =const.$ and $\mu =const.$, 
that will lead to the same equality (2.7). As a consequence from (2.8) it is 
$$\Bigl (\frac{d^2X}{dT^2},\frac{d^2Y}{dT^2},\frac{d^2Z}{dT^2}\Bigr ) = $$
$$=(1-(\lambda +2\mu )TH) 
\Bigl (\frac{d^2x}{dt^2},\frac{d^2y}{dt^2},\frac{d^2z}{dt^2}\Bigr )-
(2\lambda +\mu )H 
\Bigl (\frac{dX}{dT},\frac{dY}{dT},\frac{dZ}{dT}\Bigr ). \eqno{(2.9)}
$$
Comparing (2.7) and (2.9) we obtain $\lambda =\frac{1}{3}$ and 
$\mu =\frac{4}{3}$. Now we give the main consequences. In order to 
distinguish the influence of the time dependent gravitational potential, 
we assume Keplerian orbits for $H=0$. 

1. {From} (2.8) it follows that the quotient $\Theta_2:\Theta_1$ 
of two consecutive orbital periods observed from B is equal to 
$1+\mu \Theta H=1+\frac{4}{3}\Theta H$. This shows that 
each next orbit has a prolonged period for a factor 
$1+\frac{4}{3}\Theta H$. 
But according to the same observer B, 
the time $\Theta$ for the observer $A$ in the shielded laboratory 
is prolonged for a factor $1+\Theta H$ according to (2.5). 
Thus, according to the observer A, 
each next orbit is prolonged for the coefficient  
$$\Theta_2:\Theta_1 = \frac{1+\frac{4}{3}\Theta H}{1+\Theta H}=
1+\frac{1}{3}\Theta H.\eqno{(2.10)}$$
This is a consequence of the unholonomic coordinates. 
Formula (2.10) can be verified for the orbital periods of 
double stars (Tren\v{c}evski 2006). 
{From} (2.10) we obtain 
$$\dot{P}_b =\frac{1}{3}P_bH,\eqno{(2.11)}$$
where $P_b$ is the orbital period of the pulsar in a binary system. 
Formula (2.11) is tested for the binary pulsars 
B1885+09 (Kaspi et al. 1994) 
and B1534+12 (Stairs et al. 1998, 2002). 
which have very stable timings, 
and the results are satisfactory. Indeed, formula (2.11) together with the 
influence of decay of the orbital period caused by the gravitational 
radiation and a non-gravitational influence of kinematic nature 
in the galaxy, yield together to the measured value of $\dot{P}_b$ 
(Tren\v{c}evski 2005a, 2006). 
Notice that if we neglect the influence from the equation
(2.11), then the change of the orbital period caused by the gravitational 
radiation would not fit with the experiments. 

2. According to the observer $B$, the distance of any planet to the Sun does 
not change according to the Keplerian law, but at each moment the Keplerian 
distance to the Sun should be multiplied by the factor 
$1-\lambda HT=1-\frac{1}{3}HT$. On the other side each distance in the 
shielded laboratory is observed by B to be shortened for the coefficient 
$1-HT$. Thus, 
at each moment the Keplerian 
distance to the Sun is observed by A as multiplied by the coefficient 
$\frac{1-\lambda HT}{1-HT}=1+\frac{2}{3}HT$. Here, the distances are 
assumed to be measured via laser signals, as for the LLR. 

3. The planetary orbits are 
not axially symmetric and according to the observer B 
(Tren\v{c}evski 2006) 
the angle from the perihelion to the aphelion is 
$\pi-\frac{\lambda H\Theta\sqrt{1-e^2}}{e\pi}$=
$\pi-\frac{H\Theta\sqrt{1-e^2}}{3e\pi}$, 
while the angle from the aphelion to the perihelion is 
$\pi+\frac{\lambda H\Theta\sqrt{1-e^2}}{e\pi}$ 
=$\pi+\frac{H\Theta\sqrt{1-e^2}}{3e\pi}$,
where $\Theta$ is the orbital period and $e$ is the eccentricity 
of the orbit. This follows from (2.8). 
Indeed, this happens according to the observer B 
who observes that the distances are shorter for coefficient 
$1-\frac{1}{3}HT$. But according to the observer A, who observes that the 
distances are increasing for the coefficient $1+\frac{2}{3}HT$, the angle 
from the perihelion to the aphelion is equal to 
$\pi+\frac{2H\Theta\sqrt{1-e^2}}{3e\pi}$, while the angle 
from the aphelion to the perihelion is equal to 
$\pi-\frac{2H\Theta\sqrt{1-e^2}}{3e\pi}$. 

4. Using (2.8), or the previous discussion, the time dependent 
gravitational potential has no influence on the 
perihelion precession. 

5. In (Tren\v{c}evski 2006) are also 
considered the increasing of the orbital period of the Moon, the distance 
to the Moon and the change of the average Earth's angular velocity. 
These quantities depend on tidal dissipation and also on the time 
dependent gravitational potential. Including the influence of 
the time dependent gravitational potential, the discrepancies among 
the previous three changes become much smaller 
(Tren\v{c}evski 2006). 

\section{The Pioneer anomaly}

Now we shall see how the time dependent gravitational potential 
can explain the Pioneer anomaly. Firstly we raise some questions of such 
kind, that any viable theory that gives an explanation of the Pioneer anomaly 
should give answers. 

1. Why the considered anomaly is detected only for hyperbolic trajectories?  

2. Why the anomaly becomes observable only for heliocentric distances larger 
than 10 AU?

3. Why $a_P$ on small distances (Pioneer 11) is much smaller than its 
value on distances larger than 20 AU, where it is almost a constant?

4. Why the induced "anomaly acceleration" toward the Sun is found only for 
spacecrafts, but not for the trajectories of planets? 

We shall find the frequency $\nu =\nu_{obs}$ of the re-transmitted signal 
received by the DSN antennae. 
It will be calculated according to observers $A$ and $B$.  
Let us denote by $R$ the distance from the spacecraft to the Sun. 

The required frequency caused by the motion of the spacecraft, 
i.e. the Doppler effect, 
observed by the observer B is given by 
$$\nu =\nu_0\Bigl [1-2\frac{\frac{dR}{dT}}{c}\Bigr ],\eqno(3.1)$$
where $\nu_0$ is the initial frequency. This is just the 
Doppler redshift for the outside motion of the spacecraft. 
Here $dR$ is the increment observed from $B$, and according to 
(2.2), (2.3), and (2.4), 
$$dR=(1-tH)dr,\eqno{(3.2)}$$   
and $dT=(1+tH)dt$ according to (2.5). According to the observer $A$ 
(concerning the Doppler effect), $dt$ should be replaced by 
$dt(1+\frac{1}{3}tH)$ because of the following reason. The frequency 
$\nu_{obs}$ in (1.1) is modeled as a function of time and there is a 
dilation of the orbital time for coefficient $1+\frac{1}{3}tH$ observed
from $A$, which also has influence on $a_P$. This is equivalent 
to the statement that the 
time $dT$ in (3.1) is the "trajectory time", 
i.e. the time needed for the spacecraft to move from 
one point to a close neighboring point on the trajectory,  
which is given by 
$$dT=\Bigl ( 1+\frac{4}{3}Ht\Bigr )dt.\eqno{(3.3)}$$ 
Now, substituting the values of $dR$ and $dT$ from (3.2) and (3.3) 
into (3.1), according to the observer A (3.1) takes the form 
$$\nu =\nu_0\Bigl [1-2\frac{\frac{dr}{dt}}{c}
\Bigl (1-\frac{7}{3}Ht\Bigr )\Bigr ].\eqno(3.4)$$

Moreover, according to the observer A, in (3.4) 
will appear, also, an additional 
expression $\frac{-2HR}{c}$ 
which is analogous to the redshift observed from the 
distant galaxies. Indeed, the time which passes the light signal in two 
directions is just $\frac{2R}{c}$. Hence  
$$\nu =\nu_0\Bigl [1-2H\frac{R}{c}-2\frac{\frac{dr}{dt}}{c}
\Bigl (1-\frac{7}{3}Ht\Bigr )\Bigr ].\eqno(3.5)$$

Note that in this expression $H^2$ is neglected. 
Also, the relativistic corrections can be neglected here, because the 
unmodeled shift frequency is free of the relativistic corrections. 

Notice that the Pioneer anomaly is modeled without implementing the 
Hubble constant $H$ and also the measurements are performed according 
to the observer A, for whose point of view 
the time dependent gravitational potential is applied. 
Instead of this formula, $\nu_{obs}$ is modeled by
putting $H=0$, and according to (1.1), 
the unmodeled value $a_P$ appears. 
There, for the round-trip time period $t=\frac{2R}{c}$ of the 
signal, we obtain 
$$\nu =\nu_0 \Bigl [ 1+2a_P\frac{R}{c^2}-
2\frac{\frac{dr}{dt}}{c}\Bigr ].\eqno{(3.6)}$$

Both formulae (3.5) and (3.6) are deduced according to the observer A. 
Comparing them, we obtain 
$$a_P=cH\Bigl (\frac{7}{3}\frac{dR}{dT}\frac{T}{R}-1\Bigr ),\eqno{(3.7)}$$
which gives the required expression for $a_P$. Notice that for an arbitrary 
initial value of the time T (or t), $a_P\rightarrow \frac{4}{3}cH$. 

At this moment it is undetermined what is the initial value for $t$ (or $T$). 
This problem will be considered in the following section. 

\section {Possible values of the initial time values}

There is no global determination of the initial value of $t$, i.e. of $T$, 
because the physical laws must be independent of that choice. 
Still, we may consider different cases and make suitable choices. 
If the spacecraft moves on an elliptical orbit, 
the independence of the physical laws from the initial value 
of the time can be interpreted in the following way. Suppose that the 
orbital period of the spacecraft is $\Theta$. Then after time $t+\Theta$, 
from (3.5), we obtain 
$$\nu =\nu_0\Bigl [1-2H\frac{R}{c}-2\frac{\frac{dr}{dt}}{c}
\Bigl (1-\frac{7}{3}H(t+\Theta )\Bigr )\Bigr ].$$
Now, in comparison with (3.5), we obtain that after the period $\Theta$ when 
the spacecraft reaches the considered initial position - distance $R$ to the 
Sun with the same velocity $\frac{dr}{dt}$ at that position, 
a small blueshift will appear, such that 
$$\Delta \nu =\frac{14}{3}\frac{\frac{dr}{dt}}{c}\Theta H\nu_0.$$

But, a good interpretation for hyperbolic trajectories, without 
periodicity, is much more difficult to 
find. The interpretation in that case is given only by (3.7), but the 
biggest problem comes from the determining of $\nu_{model}$. 

Now let us calculate $R$, $T$, and 
$\frac{dR}{dt}$ as functions of one variable ($R$ or $T$). It is sufficient 
to solve the problem within Newtonian mechanics.  
If the hyperbola of motion is given by $\frac{x^2}{a^2}-\frac{u^2}{b^2}=1$,
then we choose $u$ as a parameter and obtain in final form  
$$R(u)=\sqrt{\Bigl (a\sqrt{1+\frac{u^2}{b^2}}-a\epsilon \Bigr )^2+u^2},
\eqno{(4.1)}$$
$$T(u)=\frac{a}{C}\Bigl (\epsilon u-b\cdot \ln 
\frac{u+\sqrt{u^2+b^2}}{b}\Bigr )+ \frac{K}{C},\eqno{(4.2)}$$
$$\frac{dR}{dT}(u)=u\frac{C}{a}\frac{1+\frac{a^2}{b^2}-\epsilon\frac{a^2}{b} 
\frac{1}{\sqrt{u^2+b^2}}}
{\sqrt{\Bigl (a\sqrt{1+\frac{u^2}{b^2}}-a\epsilon \Bigr )^2+u^2}
\Bigl (\epsilon -\frac{b}{\sqrt{u^2+b^2}}\Bigr )},\eqno{(4.3)}$$  
where $\epsilon$ is the eccentricity of the hyperbolic trajectory, 
$C=R^2\frac{d\varphi}{dt}=const$, and $K$ is the required constant. 
Substituting $R$, $T$ and $\frac{dR}{dT}$ from (4.1), (4.2), and (4.3) into 
(3.7) we obtain $a_P$ as function of $u$ (and hence $R$) and the 
unknown constant $K$. Notice that $a_P$ does not depend on the constant $C$
and thus we may put $C=1$. Namely, we obtain 
$$a_P=\Bigl \{\frac{7}{3}u\cdot  
\frac{\Bigl (1+\frac{a^2}{b^2}-\epsilon\frac{a^2}{b} 
\frac{1}{\sqrt{u^2+b^2}}\Bigr ) 
\Bigl (\epsilon u-b\cdot \ln \frac{u+\sqrt{u^2+b^2}}{b}\Bigr )}
{\Bigl [\Bigl (a\sqrt{1+\frac{u^2}{b^2}}-a\epsilon \Bigr )^2+u^2\Bigr ]
\Bigl (\epsilon -\frac{b}{\sqrt{u^2+b^2}}\Bigr )}-1\Bigr \} cH+ $$
$$+K\frac{7}{3}\frac{u}{a}\cdot  
\frac{1+\frac{a^2}{b^2}-\epsilon\frac{a^2}{b} 
\frac{1}{\sqrt{u^2+b^2}}} 
{\Bigl [\Bigl (a\sqrt{1+\frac{u^2}{b^2}}-a\epsilon \Bigr )^2+u^2\Bigr ]
\Bigl (\epsilon -\frac{b}{\sqrt{u^2+b^2}}\Bigr )} cH. 
\eqno{(4.4)}$$ 
Notice that $K=0$ means that measuring the time starts when the 
spacecraft reaches the perihelion position, while $K>0$ 
means that measuring the time starts 
before the spacecraft reaches the perihelion position. 
The values of $K$ may be different for Pioneer 10 and Pioneer 11. 
Moreover, $K$ may depend also on the interval where $a_P$ is considered.   
Indeed, these values are hidden in the complicated calculations of 
modeling $\nu_{model}$. Probably $K$ corresponds to a such value, 
that $a_P$ seems to be almost a constant in the chosen interval.  

The values of $a$, $b$ and eccentricity $\epsilon$ for Pioneer 10 are 
given by (Anderson et al. 2002) 

$a=1033394633$ km,\quad $b=1463395947$ km,\quad $\epsilon =1.73359601$ 

\noindent  
and the values for Pioneer 11 are given by (Anderson et al. 2002) 

$a=1218489295$ km,\quad $b=2316289260$ km,\quad $\epsilon =2.147933251$ 

The values of $u$ and 
$R$ are given in AU. Notice also that if we take 
$C=23.2609876$ for Pionner 10 and $C=33.906372$ for Pioneer 11, 
then the values of $t$ are given in years. 
The values of $a_P$ of Pioneer 10 for $K=15$   
are given on figure 1, and the values of Pioneer 11 for $K=30,45,60$ 
are given on figure 2. Notice that for Pioneer 10
$K=15$ means that measuring the time started 235.5 days before 
the spacecraft was at the perihelion.    
Notice that for this value of $K$, 
$a_P=1.18(1\pm 1.34\% )cH$ while the spacecraft was between 20 AU and 70 AU. 
Moreover, assuming that the 
mean value of $a_P$ is $8.74$ cm s$^{-2}$, then it corresponds to 
$H=76.2$ km s$^{-1}$ Mpc$^{-1}$. 
This is in accordance of the observed values of 
$a_P$, its deviations and the measured value of $H$, having in mind 
that about $10\%$ of the observed value of $a_P$ may be caused by 
another minor reasons mentioned in section 1.  
For Pioneer 11 we have a similar situation. For $K=45$ it 
means that measuring the time started 485 days before the spacecraft Pioneer 
11 was at the perihelion. For this value of $K$ we have 
$a_P=1.1725(1\pm 5.3\% )cH$. Analogously to Pioneer 10, this corresponds to  
$H=76.7$ km s$^{-1}$ Mpc$^{-1}$. 
Notice that the previous determinations
of the initial values for $T=0$, corresponds
with the time when each of Pioneer 10 and 11 
begun to follow the hyperbolic orbits. 
Indeed, after Pioneer 10 passed Jupiter and Pioneer 11 passed Saturn, 
the two spacecrafts followed escape 
hyperbolic orbits (see (Turyshev et al. 2004) and Fig. 2 there). 

Since $\frac{dR}{dt}\approx 0$ 
in case of elliptic orbits, the Pioneer anomaly 
almost disappears in this case. Moreover, $a_P$ changes its sign 
for almost circular orbits. For hyperbolic trajectories $\frac{dR}{dt}$ is 
much larger. This gives the answer to the question 1 from section 3. 

At about 9.5 AU heliocentric distance $a_P$ changes its sign 
(for K=0) and increases very quickly for Pioneer 11, 
and at about 5.5 AU heliocentric distance (for K=0) $a_P$ 
changes its sign and increases very quickly for Pioneer 10 
(but this early value of $a_P$
is not explored). This is the answer to the question 2 from section 3. 

The answer to the question 3 follows also just from the figures 1 and 2. 

Finally, the answer of the question 4 is the following. The formula (3.5) is 
deduced only for the frequencies obtained 
from the re-transmitted radio signal to the antennae, 
but not for an anomalous acceleration. We saw that the anomalous 
acceleration studied in section 2 yields to the following 
result about the change the distances to the planets. The distances to the 
planets (measured via a laser signal like the distance to the Moon) increase 
for the coefficient $1+\frac{2}{3}tH$. This change is not sufficient 
to be detected for the planets. For the distance Earth-Moon it is 
1.83 cm per year, while the measured increment is 3.8 cm per year. 
The rest part of 1.97 cm 
per year is a result of the tidal dissipation. About the other measurements
in the Earth-Moon system and the explanation according to the time 
dependent gravitational potential see (Tren\v{c}evski 2006).

\clearpage
\begin{figure}
\epsscale{1}
\plotone{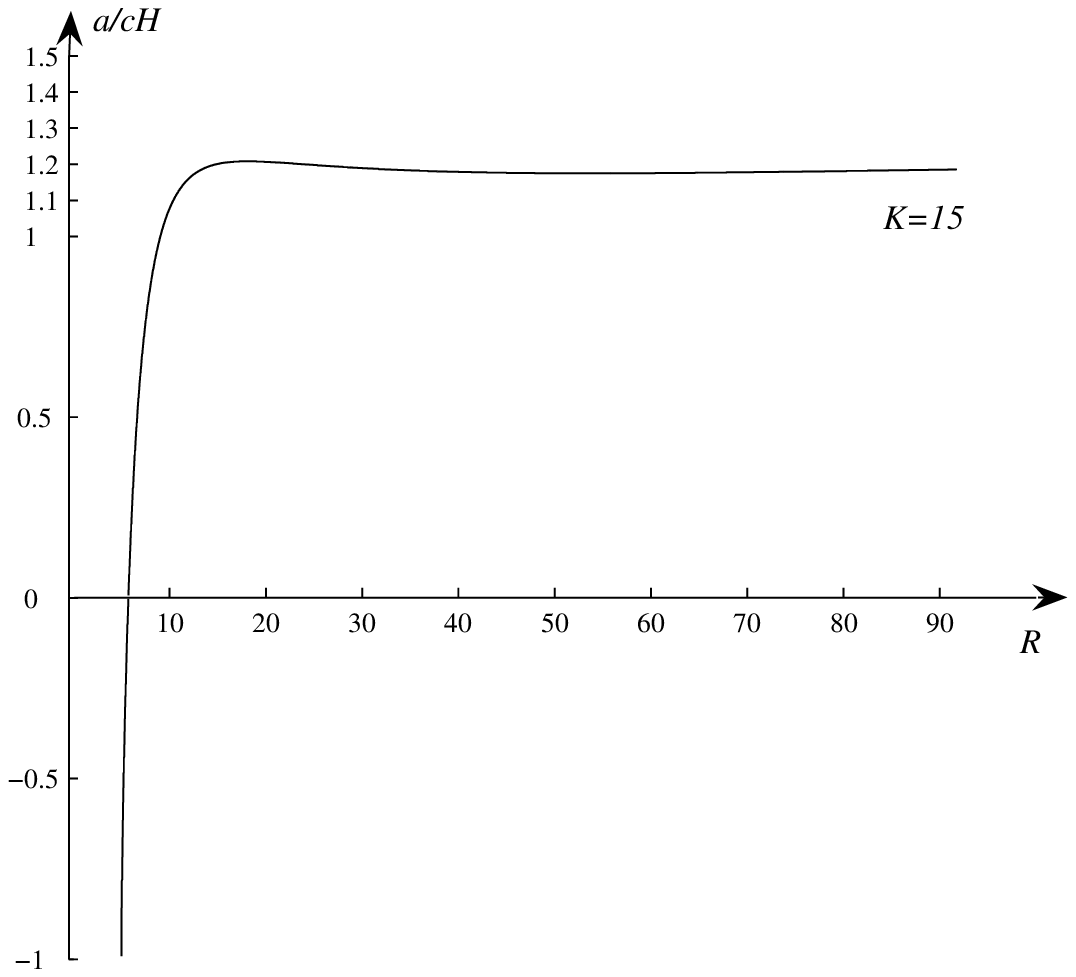}
\caption{Represents the function $a_p/cH$ for Pioneer 10 as a function of the 
distance 
R measured in AU. K=15 corresponds that measuring the time started 235.5 days 
before 
the spacecraft was at the perihelion.\label{fig1}}
\end{figure}

\clearpage
\begin{figure}
\epsscale{1}
\plotone{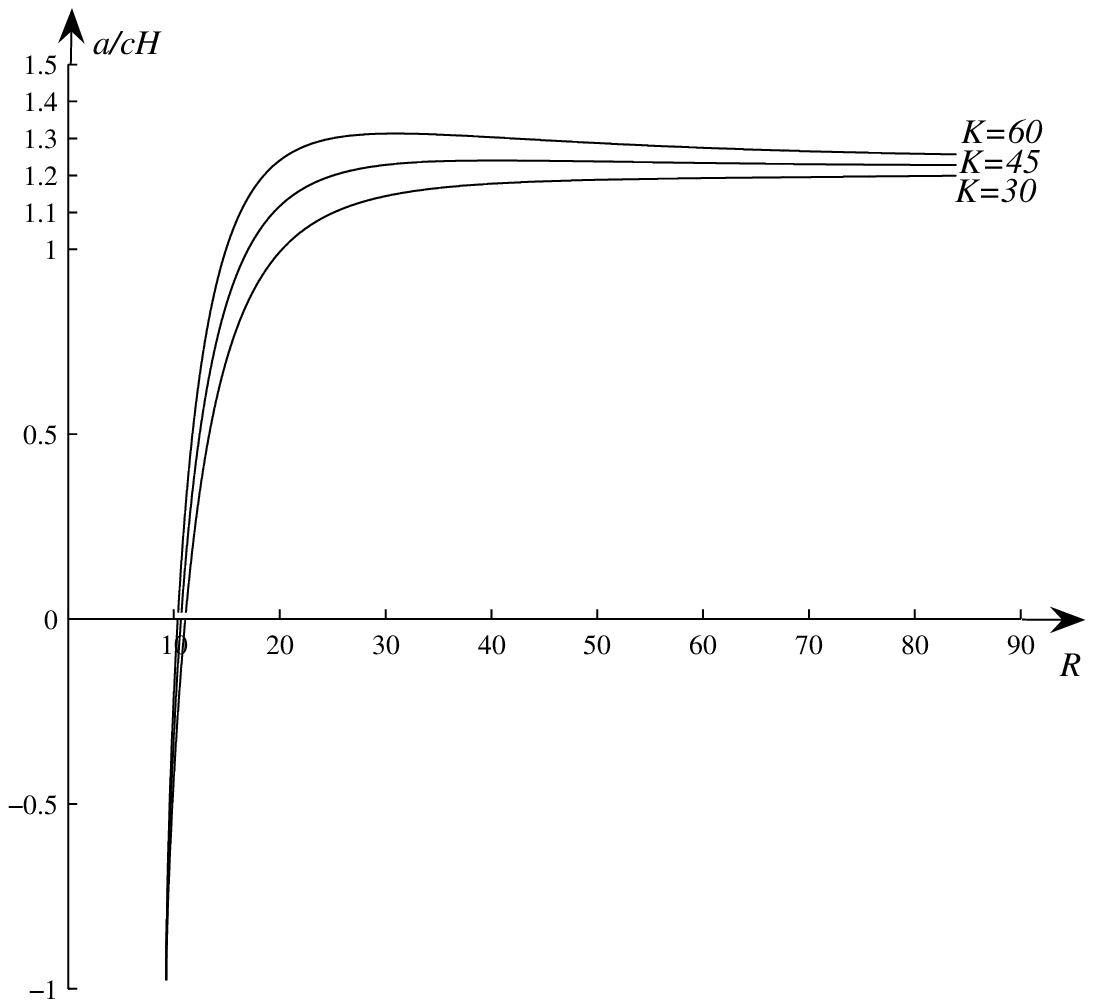}
\caption{Represents the function $a_p/cH$ for Pioneer 11 as a function of the 
distance 
R measured in AU. K=30, 45, 60 corresponds that measuring the time started 
323, 485, 647 days before 
the spacecraft was at the perihelion.\label{fig2}}
\end{figure}


\begin{thebibliography}{}

\bibitem[Anderson et al.(1998)]{And98} Anderson, J. D., Laing, P. A., Lau, E. 
L., Liu, A. S., Nieto, M. M., \& Turyshev, S. G. 1998, 
Phys. Rev. Lett., 81, 2858, eprint: gr-qc/9808081.  

\bibitem[Anderson et al.(2002)]{And02} Anderson, J. D., Laing, P. A., 
Lau, E. L., Liu, A. S., Nieto, M. M., \& Turyshev, S. G. 2002,
Phys. Rev. D, 65, 082004, eprint: gr-qc/0104064.  

\bibitem[Belayev(2001)]{Bel01} Belayev, W. B. 2001, Space Subst., 7, 63,
eprint: gr-qc/0110099.  

\bibitem[Bertolami et al.(2004)]{Ber04} Bertolami, O., \& P\'{a}ramos, J. 
2004, Class. \& Quant. Grav., 21, 3309, eprint: gr-qc/0310101. 

\bibitem[Cadoni(2004)]{Cad04} Cadoni, M. 2004, Gen. Rel. Grav., 36, 2681, 
eprint: gr-qc/0312054. 

\bibitem[Calchi et al.(2000)]{Cal00} 
Calchi Novati, S., Capozziello, S., \& Lambiase, G. 2000, Grav. Cosmol., 6, 
173 

\bibitem[Capozzielo et al.(2001)]{Cap01} 
Capozzielo, S., De Martino, S., De Siena, S., \&
Illuminati, F. 2001, Mod. Phys. Lett. A, 16, 693 

\bibitem[Consoli et al.(1999)]{Con99} Consoli, M., \& Siringo, F. 1999, 
eprint: hep-ph/9910372.  

\bibitem[Crawford(1999)]{Cra99} Crawford, D. F. 1999, Phys. Rev. Lett., 
submitted, eprint: astro-ph/9904150. 

\bibitem[Foot et al.(2001)]{Foo01} Foot, R., \& Volkas, R. R. 2001, Phys. 
Lett. B, 
517, 13, eprint: gr-qc/0108051.  

\bibitem[Guruprasad(1999,2000)]{Gur99} 
Guruprasad, V. 1999, eprint: astro-ph/9907363, 
1999, eprint: gr-qc/0005014, 2000, eprint: gr-qc/0005090. 

\bibitem[Ingersoll et al.(1999)]{Ing99}
Ingersoll, P., Johnson, T. V., Kargel, J., Kirk, R., Didon, D. I. N., 
Perchoux, J., \& Courtens, E. 1999, Preprint Universit\'{e}e de Montpellier 

\bibitem[Iorio(2006)]{Ior06} 
Iorio, L. 2006, New Astronomy, in press, eprint: gr-qc/0601055. 

\bibitem[Ivanov(2001)]{Iva01} Ivanov, M. A. 2001, Gen. Rel. Grav., 33 479,
eprint: astro-ph/0005084.

\bibitem[Kaspi et al.(1994)]{Kas94} 
Kaspi, V. M., Taylor, J. H., \& Ryba, M. F. 1994, \apj, 428, 713  

\bibitem[Mbelek et al.(1999)]{Mbe99} 
Mbelek, J. P., \& Lachi\`{e}ze-Rey, M. 1999, 
Phys. Rev. D, submitted, eprint: gr-qc/9910105. 

\bibitem[Milgrom(2001)]{Mil01} 
Milgrom, M. 2001, Acta Phys. Pol. B, 32, 3613 

\bibitem[Modanese(1999)]{Mod99} 
Modanese, G. 1999, Nucl. Phys. B, 556, 397, eprint: gr-qc/9903085. 

\bibitem[Manyaneza et al.(1999)]{Man99} 
Munyaneza, F., \& Viollier, R. D. 1999, eprint: astro-ph/9910566. 

\bibitem[Nottale(2003)]{Not03} Nottale, L. 2003, eprint: gr-qc/0307042.  

\bibitem[Ostvang(2002)]{Ost02} 
Ostvang, D. 2002, Class. \& Quant. Grav., 19, 4131, eprint: gr-qc/9910054. 

\bibitem[Ranada(2005)]{Ran05} Ra\~{n}ada, A. F. 2005, Found. Phys., 34, 1955, 
eprint: gr-qc/0403013, gr-qc/0410084.

\bibitem[Rosales et al.(1999)]{Ros99} 
Rosales, J. L., \& S\'{a}nchez-Gomez, J. L. 1999, eprint: gr-qc/9810085.

\bibitem[Sidharth(2000)]{Sid00} Sidharth, B. G. 2000, 
Nuovo Cim. B, 115, 151, eprint: astro-ph/9905052. 

\bibitem[Stairs et al.(1998)]{Sta98} 
Stairs, I. H., Arzoumanian, Z., Camilo, F., 
Lyne, A. G., Nice, D. J., Taylor, J. H., Thorsett, S. E., \&
Wolszczan, A. 1998, \apj, 505, 352, eprint: astro-ph/9712296. 

\bibitem[Stairs et al.(2002)]{Sta02} 
Stairs, I. H., Thorsett, S. E., Taylor, J. H., 
\& Wolszczan, A. 2002, \apj, 581, 501, eprint: astro-ph/0208357. 

\bibitem[Trencevski(2005a)]{Tre05a} Tren\v{c}evski, K. 2005a, 
Gen. Rel. Grav., 37 (3), 507, eprint: gr-qc/0402024, gr-qc/0403067. 

\bibitem[Trencevski(2005b)]{Tre05b} Tren\v{c}evski, K. in Int. Conf. 
on Geometry and Related Topics, Belgrade, June 26 - July 2, 2005b, 
in press 

\bibitem[Trencevski(2006)]{Tre06} Tren\v{c}evski, K. 2006, in 
Trends in Pulsar Research, ed. J.A.Lowry (New York: Nova Science), 
in press 

\bibitem[Turyshev et al.(2006)]{Tur06} 
Turyshev, S. G., Toth, V. T., Kellog, L. R., Lau, E. L., \&
Lee, K. J. 2006, Int. J. Mod. Phys. D, 15, 1, eprint: gr-qc/0512121. 

\bibitem[Turyshev et al.(2004)]{Tur04} 
Turyshev, S. G., Nieto, M. M., \& Anderson, J. D. 2004, 
in The XXII Texas Symposium on Relativistic Astrophysics, 
Stanford Univ., December 13-17, 2004, eprint: gr-qc/0503021. 

\bibitem[Turyshev et al.(2005)]{Tur05} 
Turyshev, S. G., Nieto, M. M., \& Anderson, J. D.
in XXIst IAP Colloquium on "Mass Profiles and Shopes of Cosmological 
Structures", Paris, July 4-9, 2005, eprint: gr-qc/0510081.

\bibitem[Wood et al.(2001)]{Woo01} 
Wood, J., \& Moreau, W. 2001, eprint: gr-qc/0102056. 

\end{thebibliography}
\end{document}